\documentclass[12pt]{article}
\usepackage{amsmath,amssymb,mathrsfs,url,graphicx, color}
\usepackage{cite,authblk}
\usepackage{comment}
\usepackage{placeins}
\usepackage{float}

\newcommand{\dd}{\mbox{d}}

\def\ka{\kappa}

\def\pa{\partial}

\def\ka{\kappa}
\def\ii{\textrm i}
\def\ee{\textrm e}

\def\ud{\textrm{d}}

\newcommand{\dx}{\text{d}x}

\newcommand{\id}{\textrm{d}}

\newcommand{\be}{\begin{equation}}
\newcommand{\en}{\end{equation}}
\newcommand{\bi}{\begin{itemize}}
\newcommand{\ei}{\end{itemize}}

\begin{document}

\title{Cosmic acceleration\\ from quantum Friedmann equations}

\small{
\author{Thibaut Demaerel\footnote{{Instituut voor Theoretische Fysica, KU Leuven}}, Christian Maes$^{*}$ and Ward Struyve$^{*}$\footnote{Centrum voor Logica en Analytische Wijsbegeerte, KU Leuven}}
\date{}
\maketitle

\begin{abstract}
We consider a simplified model of quantum gravity using a mini-superspace description of an isotropic and homogeneous universe with dust.  We derive the corresponding Friedmann equations for the scale factor, which now contain a dependence on the wave function. We identify wave functions for which the quantum effects lead to a period of accelerated expansion that is in agreement with the apparent evolution of our universe, without introducing a cosmological constant. 
\end{abstract}

\maketitle

\section{Introduction}

The observed acceleration in the cosmic expansion provides one of the deepest puzzles of recent cosmology. While the so-called concordance $\Lambda$CDM model agrees to a good extent with the cosmic observations, the jury is still out on the meaning and origin of its acceleration-inducing $\Lambda$-parameter \cite{copeland06,DarkEnergyReview,martin12a}. The purpose of the present paper is to present a clear and explicit case where the quantum contribution to the Friedmann equations can account for the observed expansion of the universe after the decoupling era, possibly even better than the concordance model itself (if the latter would prove insufficient \cite{verde19}).\\

Within the classical Friedmann-Lema\^itre-Robertson-Walker model of the universe the cosmological constant $\Lambda$ has traditionally been thought to provide an effective way to account for cosmological data showing acceleration in the cosmic expansion.  That model assumes a homogeneous and isotropic universe with metric
\be\label{met}
\dd s^2 = \dd t^2 - a(t)^2 \, \dd \Omega^2_k 
\en
with $a$ the scale factor and $\dd \Omega^2_k$ the spatial line element with spatial curvature $k=0,\pm 1$, for which the  Einstein equation gives rise to the Friedmann equations.  If matter is a perfect fluid with mass density $\rho$ and pressure $p$, then the scale factor enters the Hubble parameter $H= \dot{a}/a$ to satisfy
  \begin{equation}\label{1f}
  H^2 = 2 \kappa^2 \rho - k\,\frac{c^2}{a^2R^2} + \frac{\Lambda c^2}{3}
  \end{equation}
  where $\kappa^2 = 4\pi G/3$, $G$ is Newton's constant, and $R$ refers to the size of the universe. The second Friedmann equation is
   \begin{equation}\label{need}
   \frac {\ddot{a}}{a} = -\kappa^2\, \left(\rho + 3\frac{p}{c^2}\,\right) + \frac{\Lambda\,c^2}{3}
   \end{equation}
The presence of $\Lambda$ in the right-hand side thus indeed makes all the difference for cosmic acceleration, so it seems. There remain controversies on the exact nature or physical origin of that $\Lambda$. In particular, it is not settled whether the cosmological constant is strictly of quantum mechanical origin and/or whether it represents a systematic (dynamical) {\it versus} a fluctuation effect. One suggestion is that in a UV-complete theory it corresponds to the quantum vacuum energy. Another suggestion is to attribute $\Lambda$ to the back-reaction of cosmic fluctuations onto the background evolution or even to a violation of the Copernican principle by assigning a special place for the earth in the universe \cite{backreaction, Enqvist}. There have also been explorations to derive it as a quantum gravity effect \cite{copeland06,josset17,perez18,perez19}.\\
The contribution in the present paper is much more modest: it only concerns a change of physics at an effective, more macroscopic level, with quantum forces adding up to the energy-stress tensor, while local properties of the cosmos are not to be altered. We assume e.g.\ that photons still move on null geodesics and redshift according to the laws of geometric optics in general relativity.  More specifically,  we discuss the modifications of the corresponding Friedmann equations \eqref{1f}--\eqref{need} in the context of a mini-superspace description of quantum gravity following ideas of Squires \cite{squires92} and of Pinto-Neto and Santini \cite{pinto-neto03}. The set-up thus arises from applying the usual quantization techniques to a symmetry-reduced classical theory but, in contrast with \cite{pinto-neto03} where stiff matter is considered, in the present paper matter is given by dust as described by Brown and Kucha{\v r} \cite{brown95}.  We believe that is more adequate to describe the universe shortly after recombination.{\footnote{To describe the radiation dominated era, there is the Schutz formalism for general perfect fluids as described in e.g.\ \cite{schutz71,alvarenga02,letelier10,pinto-neto19}.}}  Similarly as in \cite{pinto-neto03} we find that the evolution may display accelerated expansion, without any need for an explicit cosmological constant.\\

In the case of a dust universe, the pressure term in \eqref{need} is negligible, $p/c^2 =0$, and we think of a perfect classical fluid which has very small kinetic energy compared to its rest mass.  
In the next section, we present the quantum description of gravity coupled to dust in the case of mini-superspace. The wave equation has the simple form of the free Schr\"odinger equation with the matter field playing the role of a clock variable. The quantum Friedmann equations follow from that equation by identifying the rate of change of the scale factor with the flux in the corresponding continuity equation.  We discuss the main constraints on the dynamics in Section \ref{const}.  The time-evolution of the scale factor is investigated in Section \ref{acceleration} for describing a scenario with cosmic acceleration.  We give examples of wave functions for which there is a period of accelerated expansion of the universe, without introducing a cosmological constant, yielding an excellent fit to the concordance model. We conclude in Section \ref{conclusion}.

\section{Mini-superspace with dust} \label{dust}

A possible quantum description of the coupling between gravity and matter proceeds via the Wheeler--DeWitt equation \cite{kiefer04}.  We make two simplifications. First of all, we limit ourselves to an analysis in mini-superspace by assuming homogeneity and isotropy where the wave function (of the universe) becomes a function of the scale factor and the matter field, that matter being a comoving dust (ideal pressure-less fluid). That modeling follows as a special case of the Brown-Kucha{\v r} description of dust \cite{brown95,maeda15}.\\

In the case of zero curvature $k=0$ and zero cosmological constant $\Lambda=0$, the classical description of a homogeneous Brown-Kucha{\v r} dust is determined by the Lagrangian 
\be
L = V\left[  \frac{1}{2} N a^3 \rho \left( \frac{\dot T^2}{N^2} - 1\right)  - \frac{1}{2N\ka^2}a \dot a^2 \right],
\en
where $V$ is the comoving volume, which must be such that the volume today $Va^3$ exceeds the Hubble volume \cite{maeda15}, the field $T=T(t)$ (with dimensions of time) parametrizes the four-velocity $U^\mu$ of the dust field as $U_\mu = \pa_\mu T$. $N(t)$ is the lapse function, which remains an arbritrary function that we will set to $1$ in the following. The other equations of motion then yield the equations \eqref{1f} and \eqref{need}, together with 
\be
 \frac{\ud}{\ud t}\left(a^3 \rho \right)=0  , \quad  \left(  \frac{\ud T}{\ud t}\right)^2 = 1
\en
So we get the classical dust equations together with a matter scalar $T$ that can be treated a clock variable.

The canonical quantization of this theory is straightforward and is detailed in e.g.\ \cite{maeda15}. It leads to the Wheeler--DeWitt equation{\footnote{There are operator ordering ambiguities in the canonical quantization procedure. Here, we adopt the ordering based on the Laplace-Beltrami operator corresponding to the DeWitt metric on mini-superspace \cite{christodoulakis86,maeda15}. For a discussion of other operator orderings in the context of cosmological singularities, see \cite{demaerel19b}.}}
\be\label{sch}
\ii \hbar \;\pa_T\, \psi(a,T) =  \frac{\hbar^2 \ka^2}{2Vc^2} \left( \frac {1}{\sqrt a} \pa_a \right)^2 \psi(a,T) 
\en
where $\psi$ represents the wave function of the universe in the reduced description. It is a function of the scale factor and the scalar $T$. A common way to extract predictions from this wave equation is to define the scale factor through an expectation value, i.e., $\langle a(T) \rangle = \int  |\psi(a,T)|^2 {\sqrt a}  \id a$. However, we suggest to adopt a more fine-grained approach which consists in defining the evolution of $a(t)$ and hence of the metric \eqref{met} through the streamlines of the continuity equation associated to \eqref{sch}. That is, we consider the conserved current 
\[
J = (J_a,J_T) = ( \frac{\ka^2}{Vc^2} \frac{1}{\sqrt a} \pa_a S |\psi|^2, -{\sqrt a}|\psi|^2)
\]
in terms of the polar decomposition $\psi=|\psi|\ee^{\ii S/ \hbar}$ and we put
\begin{equation}
\frac{\id}{\id t}\;
\begin{pmatrix}
a(t)  \\
T(t) 
\end{pmatrix} 
= \frac 1{\sqrt{a}\,|\psi|^2} 
\,
\begin{pmatrix}
J_a  \\
J_T
\end{pmatrix}
\end{equation}
or, explicitly,{\footnote{Those equations follow the Bohmian approach to quantum dynamics \cite{pinto-neto19}. Note that in \eqref{sch} $a$ and $T$ enter as arguments of the wave function, whereas in \eqref{bo} they are functions. The meaning will always be clear from the context.}}
\be\label{bo}
\frac{\id a}{\id t} = \frac{J_a}{{\sqrt a}|\psi|^2} \equiv \frac{\ka^2}{Vc^2} \frac{1}{a} \pa_a S , \qquad  \frac{\id T}{\id t} \equiv \frac{J_T}{{\sqrt a}|\psi|^2} =-1
\en
The last equation implies that $T = -t$ (up to a constant), like in the classical case.{\footnote{It is possible to flip the sign of $J$ (as was done e.g. in \cite{demaerel19b}), which yields the opposite velocities in \eqref{bo}.  Yet the set of solutions is the same, because they are connected by the usual time-reversal operation $\psi(a,T) \to \psi^*(a,-T)$.}} 

Introducing the constant $m=\frac{Vc^2}{\kappa^2}$ (with dimensions of mass $\times$ length$^2$), we obtain the ``quantum Friedmann equations,''
\begin{eqnarray}
\label{1fq}
&&\left(\frac{\dot a}{a}\right)^2 = \frac{1}{m^2} \frac{(\pa_a S)^2}{a^4}=:2\ka^2 \rho_{\textrm{eff}} \\
&&\frac{\ddot a}{a} = - \frac{1}{2m^2}\frac{(\pa_a S)^2}{a^4}  + \frac{F_\psi}{a}=- \ka^2 \rho_{\textrm{eff}}  +   \frac{F_\psi}{a}\label{2fq} 
\end{eqnarray}
where, 
\be F_\psi :=  a^{-1}\pa_a\left[\frac{\hbar^2}{2m^2 |\psi|}\left( \frac {1}{\sqrt a} \pa_a \right)^2 |\psi|\right]\en
may be called a quantum force. The equation \eqref{2fq} is obtained by taking the time-derivative of \eqref{1fq} and using the Wheeler--DeWitt equation \eqref{sch}.  While classical dust always decelerates the expansion, $\ddot a =-\ka^2\rho a < 0$ in  \eqref{need}, that is not necessarily so in the quantum case where $F_\psi$ may be non-zero and we may have eras of accelerated expansion as we will discover in Section \ref{acceleration}.\\ The second equality in \eqref{1fq}, i.e., the definition  of an effective density $\rho_\text{eff}$, is inspired by \eqref{1f}. Note that for a classical dust universe, we have in addition to the equations \eqref{1f}-\eqref{need} (with $p=k=\Lambda=0$) that $\id (\rho a^3)/\id t =0$, so that $\rho = \rho_0/a^3$ with $\rho_0$ a constant. The quantum dynamics \eqref{bo} reduces  to the classical one iff $(\pa_a S)^2/a$ is constant along the trajectory. But typical solutions of \eqref{sch}-\eqref{bo} do not fulfill that condition.\\ 

We continue in the next section with a general discussion on further constraints before we analyze in Section \ref{acceleration} the nature of the accelerated expansion that follows from using \eqref{sch} in \eqref{bo}.

\section{Constraints: boundary conditions and scale}\label{const}
We first simplify the dynamics and express it in terms of dimensionless variables by introducing $x:=\frac{2}{3}a^{3/2} \geq 0$ and the dimensionless time-variable $\tau :=t/t_0$ with $t_0 =  13.4$ Gyear, which is a common estimate of the age of our universe, and the dimensionless parameter $M := \frac{Vc^2}{\kappa^2t_0\hbar}=\frac{m}{t_0 \hbar}$.  The dynamics \eqref{sch} and \eqref{bo} then reduces to the simple form
\begin{eqnarray}
&& \ii \,\pa_\tau \psi  =  -\frac{1}{2M}\,\pa^2_x \psi\label{bo3}\\
&& \frac{\id x}{\id \tau} = \frac{1}{M}\,\pa_x s, \qquad \psi=|\psi|\ee^{\ii s}\label{guide2}
\end{eqnarray}
which are the familiar equations for a non-relativistic free particle on the half-line with mass $M$.  

In terms of these variables, the quantum Friedmann equation \eqref{2fq} reads
\be
M \frac{\id^2 x}{\id \tau^2} = {\tilde F}_\psi , \qquad   {\tilde F}_\psi = \frac{1}{2M}\pa_x  \frac{\pa^2_x |\psi|}{|\psi|} 
\en
Classical evolution is obtained when ${\tilde F}_\psi$ is zero. This will be the case for plane wave solutions to the Schr\"odinger equation \eqref{bo3}. For a generic wave function, there will be a deviation from classicality and hence a potential effective cosmological constant. For square integrable wave functions (over $x \in {\mathbb R}$, with \eqref{bo3} then viewed as the restriction to the positive half-line --- more on this in the next section), classicality will be obained for large $x$ and $t$ \cite{duerr09}, due to the dispersive character of the free Schr\"odinger evolution. So in the case of these wave functions, the appearance of an effective cosmological constant can merely be a transient feature.

\subsection{Boundary conditions}
As usual, the dynamics \eqref{bo3} may be taken as unitary on the Hilbert space L$^2(0,+\infty)$ with flux-prohibiting boundary conditions: $\psi|_{x=0} = \ell\, \pa_x \psi|_{x=0}$ with $\ell \in \mathbb{R} \cup \{\infty\}$, see e.g.\ \cite{maeda15}. These boundary conditions imply that the density $|\psi(x,t)|^2$ is preserved by the dynamics \eqref{guide2} and can be taken as a probability distribution. Moreover, since $(|\psi|^2\pa_x S)|_{x=0} = 0$, trajectories never visit $x=a=0$ (after a finite cosmic proper time). In other words, there is never a big bang or big crunch. Trajectories can also not reach $+\infty$ in finite time, which means that there is no big rip. 

Despite appearances however, the dynamics \eqref{bo3} is not quite the usual Schr\"odinger equation; it remains a Wheeler-deWitt equation in which the variable $\tau$ started out to represent the dust content. It is then less clear what is the Hilbert space or why the wave function would satisfy such a boundary condition as above. Rather we may consider and allow a flux through $x=0$ which leads to the possibility of a big bang or big crunch. For example, the Vilenkin-type wave functions are of such a kind \cite{vilenkin82}.\\

In the following section, we consider wave functions for which there is flux through $x=0$ as well as a wave function for which there is no flux. The former will give big bang trajectories while the latter will give bouncing trajectories (where a minimal spatial volume is reached). In any case, we must remember that we merely consider a dust model of our universe, and near the big bang the trajectories should not be taken too seriously as that concerns a radiation dominated era.

\subsection{Scale}\label{sca}
The parameter $M$ in the equations \eqref{bo3}--\eqref{guide2} depends on $V$. To model our universe, the spatial volume should at least be our current Hubble volume, which is of the order $(10^{10}$lightyear$)^3$. Fixing the value of the scale factor $a$ today to be one, we take $V=(10^{10}$lightyear$)^3$. As such, $M$ is roughly of the order of $10^{120}$.\\

Because of our choice of coordinates $x$ and $\tau$, we will aim to fit the evolution of the universe roughly between $(x,\tau)$ equal to $(0,0)$ and $(2/3,1)$.  Therefore the expansion rate must be of order one. Looking at \eqref{guide2}, since the ``mass'' $M$ is exceedingly large we must have $\pa_x S$ large as well. That can be obtained in various ways but it does constrain the possible wave functions or initial conditions. As an example, consider a Gaussian wave function,
\begin{eqnarray}
&& \psi_{{\bar x},v,\sigma}(x,\tau) = \sqrt{\frac{\sigma}{ \sqrt{\pi} (\sigma^2 - \ii \tau/M ) }}  \ee^{\ii M(v x -  v^2 \tau/2)} \ee^{ - \frac{(x - {\bar x} - v\tau)^2}{ 2 (\sigma^2 + \ii \tau/M)  }}
\label{G1} \\
&& x(\tau) =  {\bar x} + v\tau + [x(0) -  {\bar x} ] \sqrt{1 + \frac{\tau^2}{\sigma^4M^2}} \label{G2}
\end{eqnarray}
To have the desired expansion rate, we take $v$ of order 1 and $\sigma$ such that $\sigma^2M \gg 1$. Then, the wave packet has a high ${\cal O}(M)$ momentum. (Alternatively, we could take $\sigma$ appropriately small, but then $x(0)$ would have to be far away from the bulk of the packet.) 

A second problem related to the large value of $M$ is numerical.  To solve \eqref{bo3}--\eqref{guide2} with parameter $M=10^{120}$ requires the time-steps to have a size of order $10^{-120}$. As shown in the Appendix, that problem can be overcome by instead solving directly the limiting case $M \to \infty$. By taking this limit a simplified integration method is available.

\section{Acceleration as a quantum effect}\label{acceleration}
As is apparent from the second quantum Friedmann equation \eqref{2fq}, it is the quantum force $F_{\psi}$ that enables acceleration of the expansion. Of course, not all (initial) wave functions $\psi$ lead to universes compatible with cosmological observations. Here we investigate how easy it is to get full qualitative and even quantitative compatibility for some class of wave functions.\\ We first consider three different wave functions: two different superpositions of Gaussian wave packets for which the evolution of the scale factor is obtained numerically and one for an Airy wave where the solution is exact. For each case we aim to fit a $\Lambda$CDM universe with parameters $H_0=70$ km/s/Mpc (the Hubble parameter today), $\Omega_M=1-\Omega_{\Lambda}=0.3$ (the matter and dark energy density parameter): hereafter this model is referred to as the ``standard $\Lambda$CDM model". More precisely, the present day scale factor and Hubble rate are fitted exactly, while the evolution of the modeled universe may deviate from the one corresponding to the $\Lambda$CDM universe near the big bang era (in particular the big bang might be replaced by a bounce), since the experimental data are less restrictive there.  We end the section by a discussion on the {\it genericity} of wave functions leading to the observed cosmic acceleration.

\subsection{Gaussian superposition with big bang trajectories}\label{2Ga}
Using the notation \eqref{G1}, we consider first a superposition of two Gaussian wave functions, 
\begin{equation}\label{g2}
\psi(x,\tau)=\alpha_1\,\psi_{\bar{x}_1,v_1,\sigma_1}(x,\tau)+\alpha_2\,\psi_{\bar{x}_2,v_2,\sigma_2}(x,\tau)
\end{equation}
with parameters
\begin{equation}\label{2g}\bar{x}_1=0,\quad\bar{x}_2=-1,\quad v_1=0.3,\quad v_2=2,\quad \sigma_1=0.35,\quad \sigma_2=1,\quad \alpha_1^2=1-\alpha_2^2=0.59\end{equation}
This wave function carries incoming flux at the origin $x=0$: all trajectories $x(\tau)$ move from the negative reals at early times to the positive reals later on. When passing through $x=0$ a big bang occurs whereby the universe is newly created.  The two packets move at high but comparable momentum towards larger $x$ with the second one overtaking the first one, at which moment the acceleration sets in. We refer to the Appendix for the numerical recipe that produces the trajectories.\\
In all, we get a very good fit to our $\Lambda$CDM universe up to a redshift $z=\frac{1}{a}-1$ of about 4 (corresponding to the time interval $\tau \in [0.2,\,1]$); see Fig.~\ref{2Gauss}. This model predicts that, in the far future, the cosmic expansion will decelerate again and converge onto a trajectory of the form $a(\tau)\sim (3(\tau-\tau_0))^{2/3}$, as exhibited in the lower part of Fig.~\ref{2Gauss}.  Moreover  the acceleration remains relatively mild without excessive choices of the parameters \eqref{2g} (the velocities and widths  are of the same order in the two packets). Therefore, that those trajectories in bounded time-intervals resemble ``coincident" $\Lambda$CDM models (i.e., those with $\frac{\Omega_{\Lambda}}{\Omega_{\text{dust}}}=\frac{\rho_{\Lambda}}{\rho_{\text{dust}}}={\cal O}(1)$, \cite{Velten})  seems to appear here ``naturally.''

\begin{figure}[h]
\centering
\includegraphics[width=\textwidth]{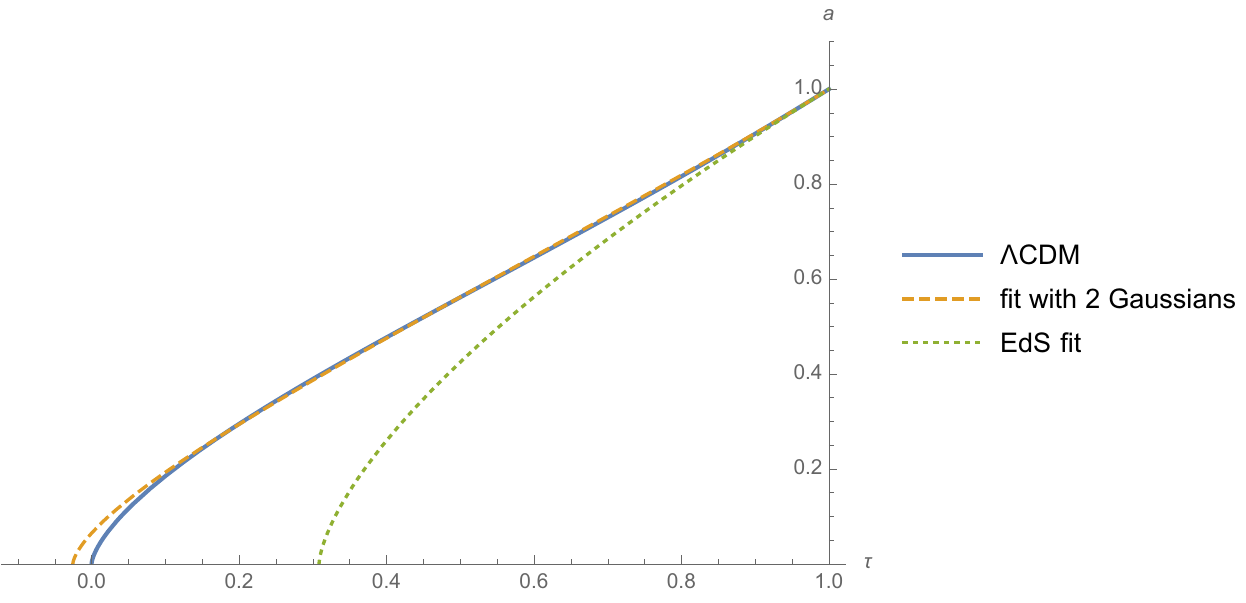}
\includegraphics[width=\textwidth]{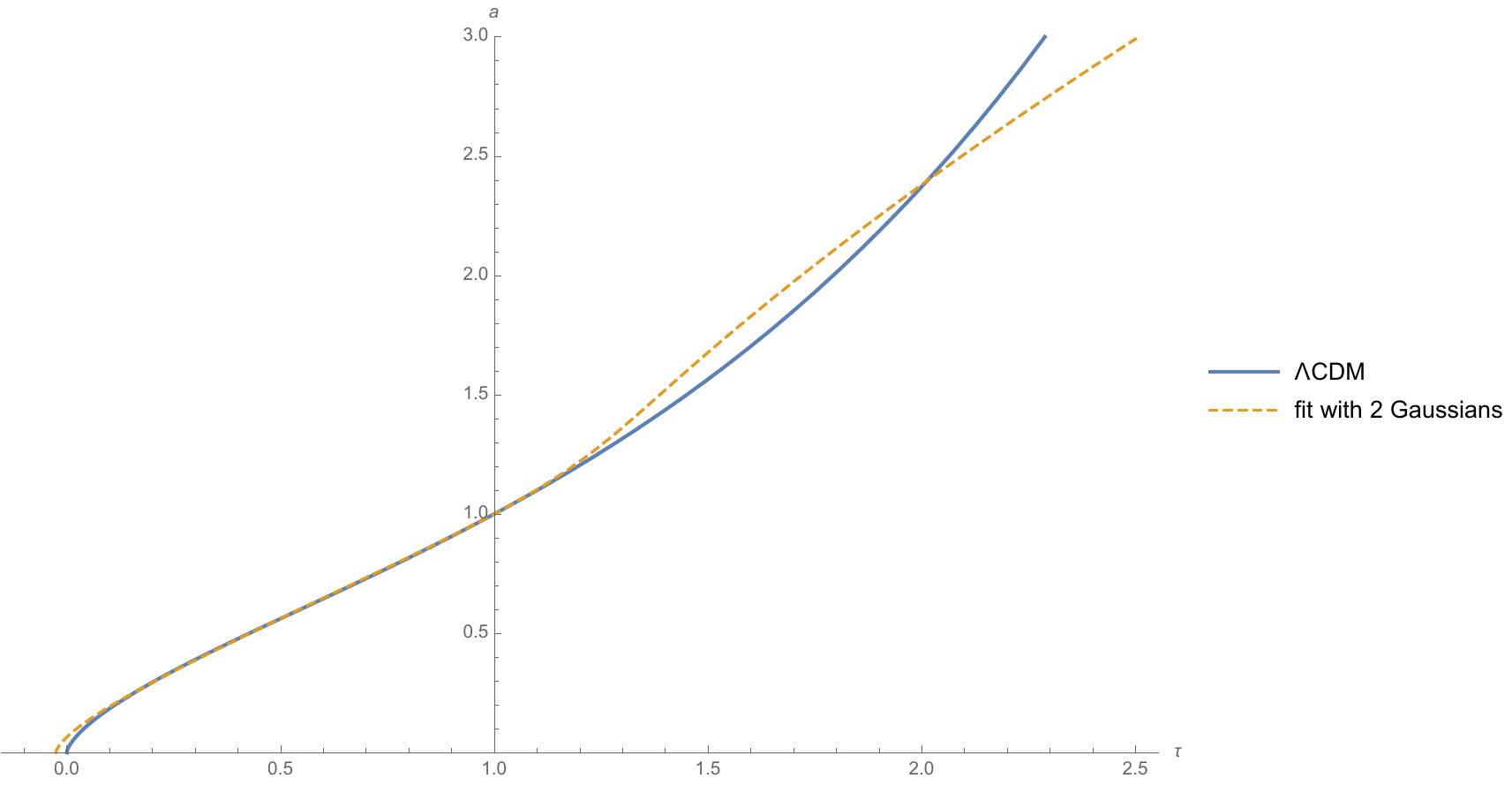}
\caption{The evolution of the scale factor $a(\tau)$ in the case of the Gaussian superposition \eqref{2g}. Upper plot: the current history of the standard $\Lambda$CDM model and a fitting trajectory (which agrees very well). We also compare with an Einstein-de Sitter universe, i.e., with a dust-dominated universe, $a(\tau)\propto (\tau-\tau_B)^{2/3}$ fitting the observed expansion $H_0$ of today. Lower plot: a longer-time comparison of the evolution of the standard $\Lambda$CDM model with the same trajectory, showing a deviation in our future.}
\label{2Gauss}
\end{figure}
\FloatBarrier

\subsection{Gaussian superposition with bouncing trajectories}\label{SGauss}
In this Section we consider a unitary evolution for the following superposition of four Gaussians:
\begin{equation}\label{4g}
\psi(x,\tau)=\frac{5}{\sqrt{74}}[\psi_{0,\,v_1,\,\sigma_1}(x,\tau-\tau_0)+\psi_{0,\,-v_1,\,\sigma_1}(x,\tau-\tau_0)]+\frac{7}{\sqrt{74}}[\psi_{0,\,v_2,\,\sigma_2}(x,\tau-\tau_0)+\psi_{0,\,-v_2,\,\sigma_2}(x,\tau-\tau_0)]
\end{equation}
where $v_1=0.733$, $v_2=0.975$, $\sigma_1 = 5.64\times 10^{-6}$, $\sigma_2=0.705$, $\tau_0=3.84\times 10^{-2}$. This wave function satisfies the Neumann boundary condition towards $x=0+$ and hence its L$^2$-norm is preserved on the half-line $(0,+\infty)$. As said before, that implies that trajectories never start with a big bang or end up in a big crunch. The value of $\sigma_1$ is chosen such that there is a bounce with a minimal scale factor within $10^{-4}$ for the range of initial conditions we are interested in: in our universe that scale factor lies beyond the surface of last-scattering and well in the radiation-dominated epoch.

\begin{figure}[h!]
\centering
\includegraphics[width=\textwidth]{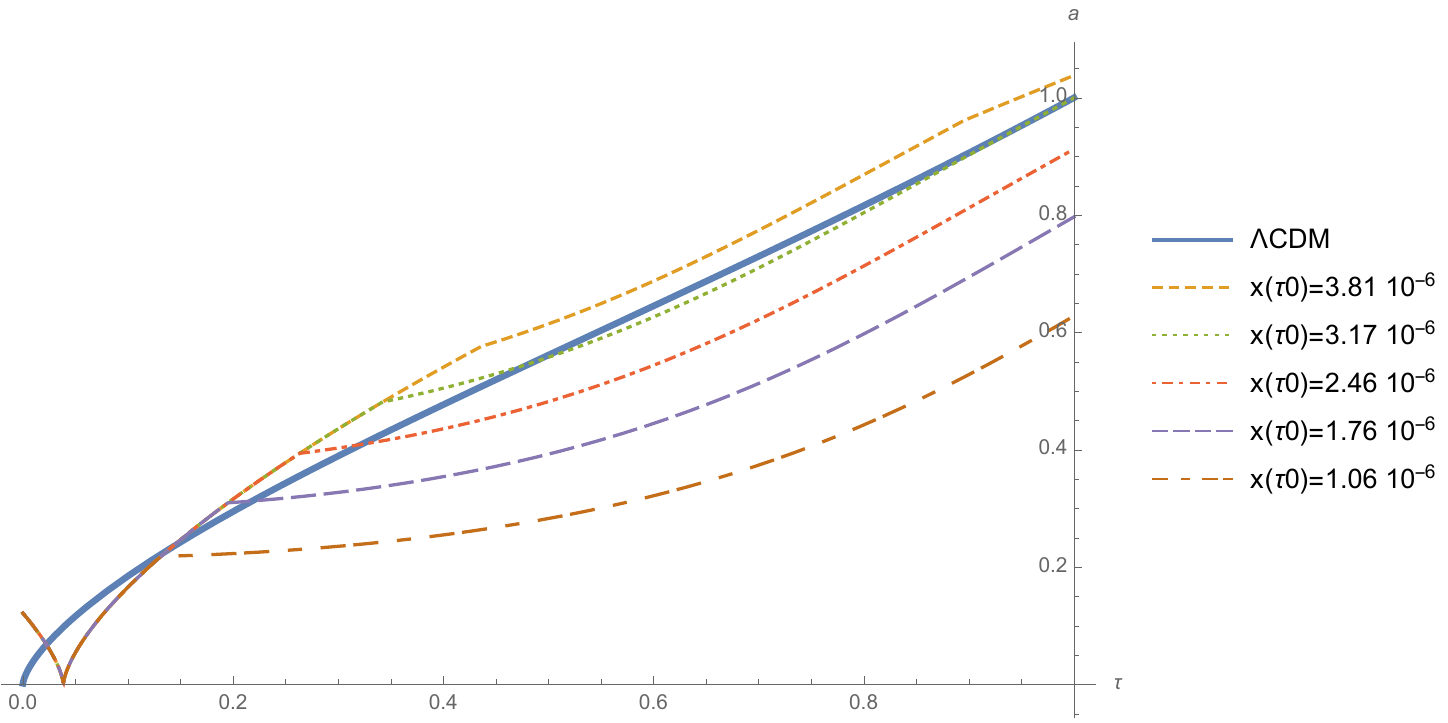}
\caption{The scale factor $a(\tau)$ for the Gaussian superposition \eqref{4g}. The bold blue curve corresponds to the standard $\Lambda$CDM model while the other curves are ordered top-down in the same order as in the legend. The best fit to the $\Lambda$CDM model is given by the trajectory with initial $x(0)=3.6\times 10^{-6}$. Other initial scales lead to trajectories with overall the same qualitative features. Note that the big bang that is implied by the $\Lambda$CDM model is replaced by a bounce (which despite appearances is smooth).}
\label{Gauss}
\end{figure}
\FloatBarrier
To numerically calculate the trajectories, we again use the limit $M \to \infty$ as explained in the Appendix. In Fig.~\ref{Gauss}, we consider 5 different initial values $x(0)\in I := [1.2\times10^{-6},\,4.4\times10^{-6}]$. For each of those values,  the scale factor $a(t)$ decelerates (the abrupt change in speed stems from the fact that $\sigma_1 \ll 1$) and then accelerates before finally (in our future) to decelerate again onto a classical $a(\tau)\propto\tau^{2/3}$ trajectory.  The first epochs are in qualitative agreement with the $\Lambda$CDM model. The $|\psi|^2$-probability of the range $I$ is about $15\%$, so that these initial conditions are not particularly special.

\subsection{Airy wave train}
Finally, we consider the Airy packet solution, \cite{berry79},
\be\label{A}
\psi(x,\tau) = {\textrm{Ai}}\left(M^{2/3}B\left[x-B^3(\tau-\tau_0)^2/4\right]\right) \exp\left\{\ii M \frac{B^3(\tau-\tau_0) [x-B^3(\tau-\tau_0)^2/6]}{2}\right\}
\en
parameterized by $B$ and $\tau_0$.  Here, \cite{holland93b},
\be\label{A1}
x(\tau)=\frac{2}{3}a^{\frac{3}{2}}(\tau) = \frac{B^3(\tau-\tau_0)^2}{4} + x(\tau_0)
\en

For late times, $a(\tau) \sim \tau^{4/3}$, which is the classical motion corresponding to a perfect fluid with equation of state  $p=-\rho/2$. That illustrates that quantum solutions may behave very differently from the classical ones.\\

 If $B>0$ and $x(\tau_0)\geq 0$ (which implies that $x(0) >0$), then the trajectory corresponds to a bouncing universe. (So even though there is a non-zero flux through $x=0$, a particular trajectory may never reach $x=0$). Moreover, such trajectories are always accelerating, very much unlike classical trajectories which are constantly decelerating. Unlike the examples in the preceding section, there is no classical evolution for large $x$ and $t$, because the Airy wave train is a non-dispersive (and non-normalizable) packet when viewed over the whole line $x \in {\mathbb R}$.

\begin{figure}[h!]
\includegraphics[width=\textwidth]{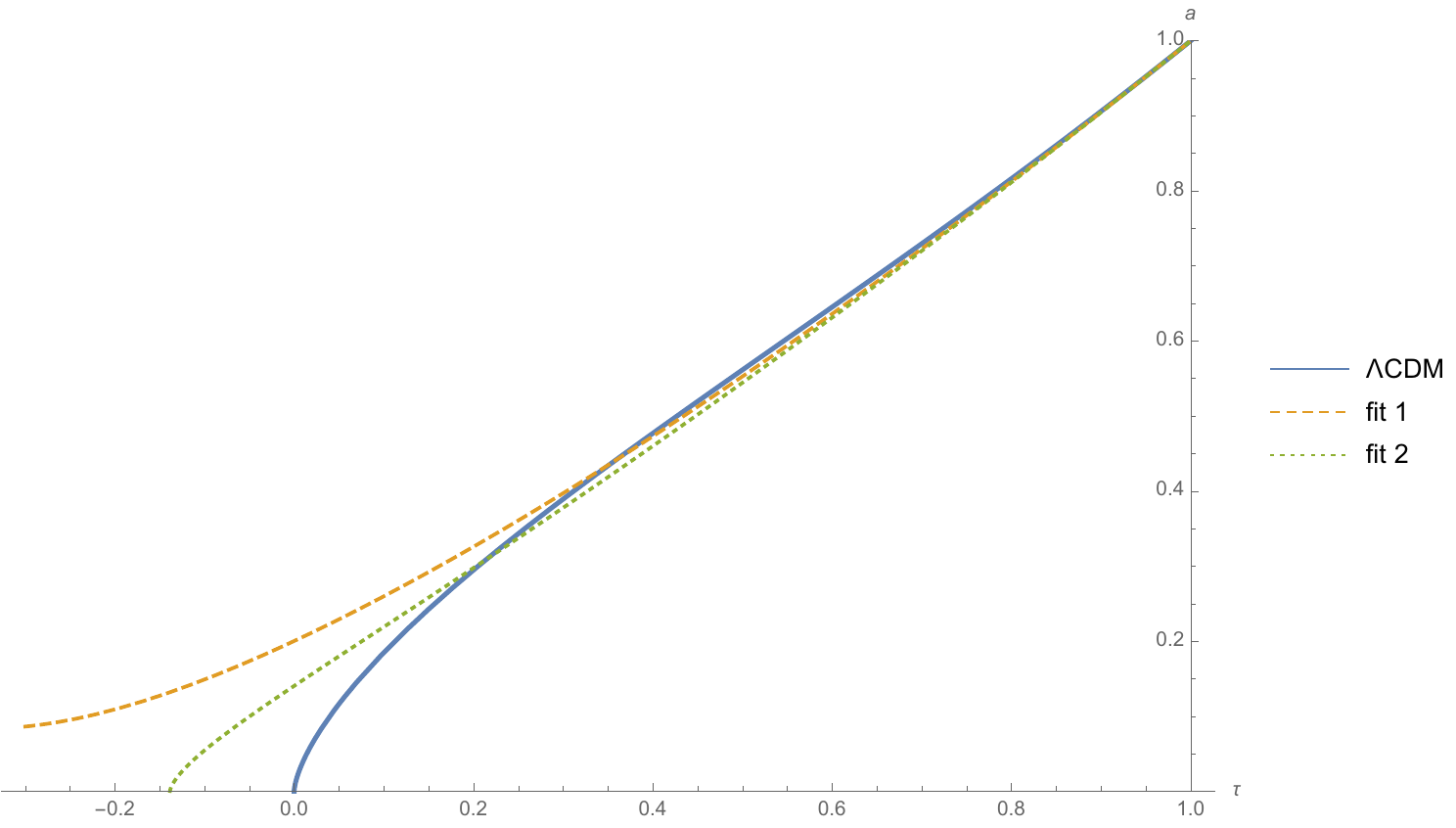}
\caption{The scale factor $a(\tau)$ fitting the standard $\Lambda$CDM model for the Airy wave train \eqref{A}.}
\label{Airyf1}
\end{figure}
\FloatBarrier

On the other hand, when $B>0$ and $x(\tau_0)<0$ (for which $x(0)=0$ is possible), there is a big bang with an initial period of deceleration. Fits of trajectories of the latter type to the standard $\Lambda$CDM model are displayed in Figs.~\ref{Airyf1} and \ref{Airyf2}. It is clear that $x(\tau_0)$, $\tau_0$ and $B$ can be tuned so that e.g.\ the present-day Hubble parameter $H_0$ is fitted. All considered fits do just that. The remaining free parameter, $\tau_0$, equals $-0.35$ for fit 1 and $-0.45$ for fit 2.

\begin{figure}[h!]
	\includegraphics[width=\textwidth]{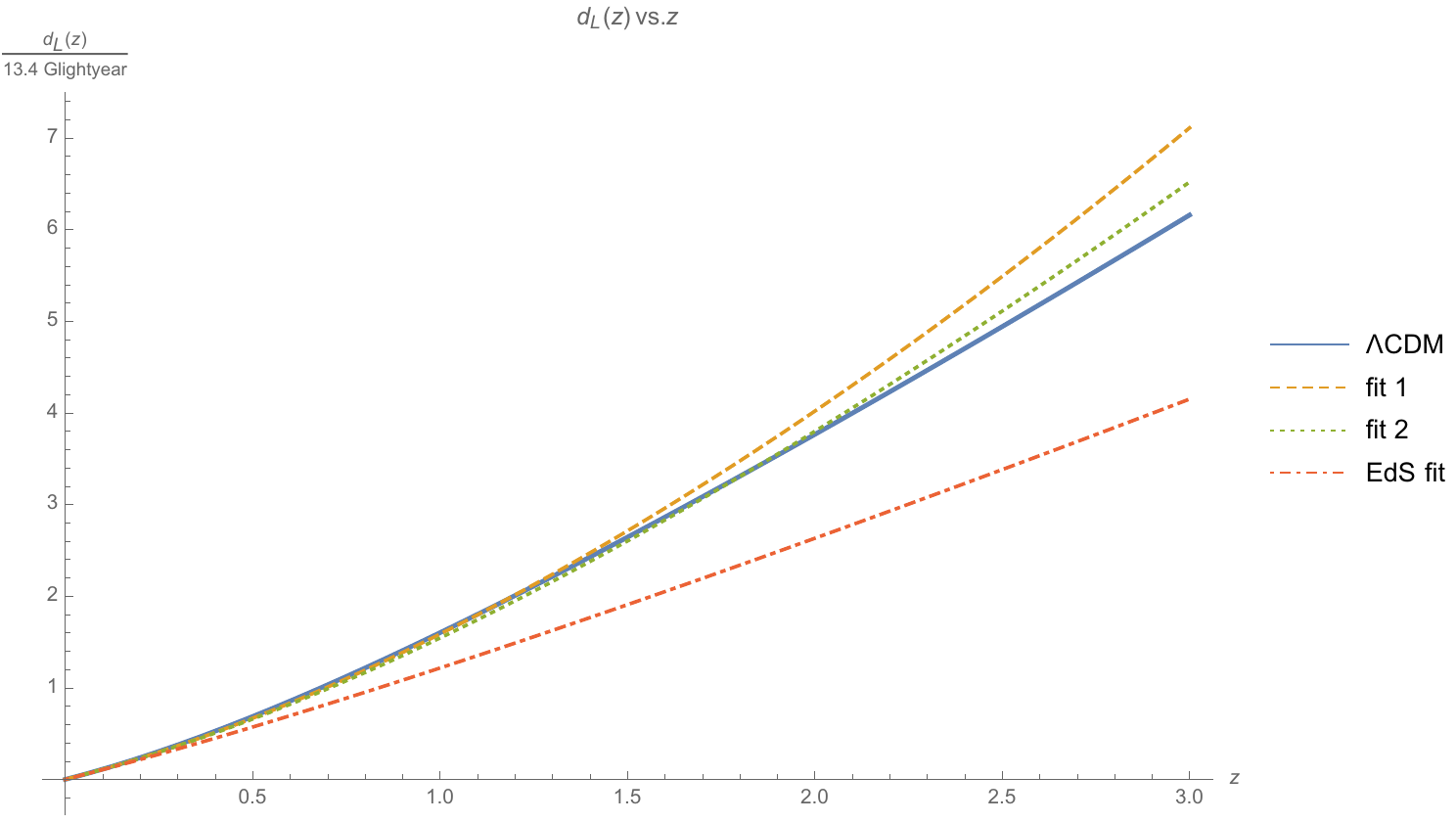}
\caption{The luminosity distance $d_L(z)$ {\it versus} redshift $z=\frac{1}{a}-1$, fitting the standard $\Lambda$CDM model for the Airy wave train \eqref{A}.  The lower curve corresponds to an Einstein-de Sitter universe, i.e.,  dust-dominated, $a(t)\propto t^{2/3}$, and fitting the expansion $H_0$ of today.}
\label{Airyf2}
\end{figure}
\FloatBarrier

\subsection{Genericity, decoherence and optics}
One may wonder how generic and under what conditions a period of accelerated expansion arises. The evolution of the scale factor is determined by the initial wave function and initial value of the scale factor. We have seen that with a rather simple wave function (i.e., a superposition of two Gaussians), we have the desired period of accelerated expansion. And for those wave functions, such a period is rather generic. Of course, there are also other wave functions, which don't give the desired behaviour. As already noted, the plane waves give classical motion. Other wave functions will give consecutive periods of acceleration (e.g., superpositions of a higher number of Gaussians). While it may be interesting to quantify that in great detail, one should remember that as a matter of fact, we only need ``one good wave function,'' in the same sense as we only have ``one universe.''

A second point of discussion concerns the classical versus quantum nature of the description. Does quantum physics not only matter for early universe considerations? The majority position is probably 'yes.'   Our work does give quantum corrections to the motion of the scale factor in recent times, and that is clearly questionable. Therefore further work is required to show that the result still holds when we go beyond the mini-superspace approximation. It will then be relevant to consider possible decoherence effects that might arise from the coupling between homogeneous and inhomogeneous degrees of freedom. This  decoherence in the quantum evolution of the universe is not exactly a well-understood phenomenon.  Estimates of decoherence times for the scale factor are difficult, also in the many other approaches to the dark energy problem.  At least, in the present work, we see precisely that the  very long-time solutions of our quantum Friedmann equations become indeed classical: the geometry of the late future Universe is described by a classical solution of Einstein's equations.

Finally, we stress that our study did not concern local properties of the cosmos. We gave predictions for specific trajectories of the scale factor. That ‘only’ gives a motivated correction to the evolution of the scale factor (i.e., a quantum modification in the FLWR-metric for a dust universe); the laws of free-fall motion in the resulting curved space are the same as always.

\section{Conclusion}\label{conclusion}
By quantizing a symmetry-reduced classical theory with an effective description of a dust fluid in terms of a non-canonical scalar field, we have obtained the quantum analogue of a Friedmann-Lema\^itre-Robertson-Walker universe with dust as matter. The quantum treatment of such a mini-superspace model readily allows for deviations compared to the classical motion. These deviations can be understood in terms of a ``quantum force'' appearing in the second quantum Friedmann equation. We have identified solutions for which there is a period of accelerated expansion which matches the classical evolution of the standard $\Lambda$CDM model very well for the period of the age of the universe.  There are also differences. Namely, the effective cosmological constant is usually transient. Other solutions could also lead to, say, consecutive eras of accelerated expansion. This fact is interesting in the light of the recent puzzle concerning the discrepancy in the measured values of the Hubble rate, measured on the one hand by the Hubble Space Telescope and by the Planck satellite \cite{verde19}. This discrepancy, which does not appear to be resulting from a systematic error, could for example be explained by the appearance of an effective cosmological constant at early times \cite{poulin19}. While this was modelled by a scalar field in \cite{poulin19}, it could also appear naturally in our model without having to introduce a scalar field.\\
While our analysis by technical necessity has been limited to a simplified set-up, the immediate success still suggests that the mechanisms discussed in the paper will also be at work in full quantum gravity. The simplicity by which we obtain dark energy could also be contrasted by the difficulties to get it from supersymmetry or string theory.

\section{Acknowledgments}
It is a pleasure to thank Patrick Peter for valuable comments. WS is supported by the Research Foundation Flanders (Fonds Wetenschappelijk Onderzoek, FWO), Grant No. G066918N.

\appendix

\section{Appendix: the $M \to \infty$ limit}
Because $M$ is large, the numerical analysis of the evolution of the scale factor is problematic. We can overcome this problem by considering the limit $M \to \infty$.
 
Before we consider this limit, note that for a solution $\{x(\tau),\psi(x,\tau)\}$ to \eqref{bo3} and \eqref{guide2}, we have that \cite{goldstein99}:
\begin{equation}\label{equi}
 x(\tau)\equiv P_{\tau}^{-1}(P_{0}(x(0)))
\end{equation}
where $P_\tau(x)$ is the cumulative probability distribution
\[P_{\tau}(x):=\int_x^\infty |\psi(x',\tau)|^2\dx'\] 
That is, trajectories correspond to the contours of the cumulative distribution function. 

Consider now a Gaussian superposition (normalized over the real line)
\begin{equation}\label{supG}
\psi_M(x,\tau) = \sum_{j=1}^N \alpha_j \psi_{{\bar x}_j,v_j,\sigma_j,M}
\end{equation}
It can be shown that the corresponding trajectories $x_M(\tau)$ of the rescaled scale factor converge uniformly to $x_{\infty}(\tau)$ under the limit $M \to \infty$ (with $\overline{x}_j$, $\sigma_j$, $v_j$ constant), where $x_{\infty}(\tau)$ is the contour of the following cumulative probability distribution
\begin{multline}\label{limM}
P_{\tau,\infty}(x):=\lim_{M\to \infty}\int_x^\infty |\psi_M(x',\tau)|^2\dx'=\sum_{j=1}^N |\alpha_j|^2\lim_{M\to \infty}\int_x^\infty |\psi_{{\bar x}_j,v_j,\sigma_j,M}(x',\tau)|^2\dx'\\
 =\sum_{j=1}^N |\alpha_j|^2 \int_x^\infty\frac{1}{\sigma_j\sqrt{\pi}} e^{-\frac{(x'-\bar{x}_j-v_j\tau)^2}{\sigma_j^2}}\dx'
\end{multline}
i.e.,
\begin{equation}\label{limit traj}
x_{\infty}(\tau)=P_{\tau,\infty}^{-1}(P_{0,\infty}(x_{\infty}(0)))
\end{equation}

Assuming for simplicity that all the velocities $v_j$ in \eqref{supG} are different and letting $P_{\tau,M}$ be the cumulative distribution function associated to the wave function \eqref{supG}, we have that $P_{\tau,M}$ converges to \eqref{limM}, in the sense that $\|P_{\tau,M}-P_{\tau,\infty}\|_{\infty}\leq \frac{C}{M}\sup_{1\leq i<j\leq N}\frac{1}{|v_i-v_j|}$ for some parameter-independent ${\cal O}(1)$ constant $C$. Finally, it can be shown that a converging $P_M$ gives rise to converging orbits $x_M$ as $M\uparrow \infty$.\\
  In conclusion, the error in numerically solving the trajectories made by replacing $M=10^{120}$ with $M=\infty$ is indeed negligible, and we can use the far simpler recipe provided by \eqref{limit traj}.

\end{document}